\documentstyle[psfig,aasms4,12pt]{article}
\newcommand{\bea}{\begin{eqnarray}} %%not used
\newcommand{\be}{ \begin{equation}}
\newcommand{\eea}{\end{eqnarray}}    %%not used
\newcommand{\ee}{\end{equation}}
\newcommand{\bc}{ \begin{center}}
\newcommand{\ec}{\end{center}}
\newcommand{\bi}{ \begin{itemize}}
\newcommand{\ei}{\end{itemize}}
\newcommand{\beno}{ \begin{eqnarray*}}

\newcommand{\eten}{\eta_{10}}
\newcommand{\Ombe}{\Omega_B h^2}  
\newcommand{\Om}{\Omega_{cl}}
\newcommand{\Omm}{\Omega_m}

\newlength{\mswsize}
\begin{document}

\begin{center}
{\Large {\bf Baryonic Mass Fraction in Rich Clusters and the Total
		 Mass Density in the Cosmos}}
\end{center}
 
\begin{center}
{\bf \large S.A. Bludman} \\
{\em 
Department of Physics, University of Pennsylvania,               \\              
Philadelphia, Pennsylvania 19104 \\}

%%%%%**preprint**
               
      (August 28, 1997, UPR-694T, astro-ph/9706047) \\         % in for **preprint**

\vspace{0.3cm}

\end{center}

\begin{abstract}

Direct observations of the supposedly universal primordial deuterium
abundance imply a relatively large baryon density $\Omega_B=
(0.019-0.030)h^{-2}$ (95\% C.L.).  On the other hand, concordance
between the previously accepted $^4 He$ and $^7 Li$ abundances and
standard Big Bang Nucleosynthesis requires the thrice smaller value
$\Omega_B=(0.005-0.010)h^{-2}$ (95\% C.L.).  For each $\Omega_B$, we use
X-ray and Sunyaev-Zeldovich observations of the baryon fraction $f_B$
in rich clusters of galaxies, in order to obtain limits on the total
mass density $\Omega_{cl}$ in clusters of galaxies.  The
higher-$\Omega_B$ values are consistent with clusters being a fair
sample of the Universe, and then imply $\Omega_{m} =(0.3-0.9)$, a
medium or critical density Universe.  Said otherwise,
the observed limits $f_B > 0.1, ~
\Omega_m > 0.3$ imply $\Omega_B > 0.03$. If the newer$^4 He$ abundance 
observations are accepted, this is consistent with standard BBN.
\end{abstract}

%\pacs{PACS numbers: }

%\setlength{\baselineskip}{2.6ex}
%
\section{DIFFERENT BARYON/PHOTON RATIOS OBTAINED FROM BBN}

\subsection{The Former Crisis in Standard Big Bang Nucleosynthesis}

Deuterium has been traditionally used as a baryometer.  Until
recently, observations of the chemically evolved nearby interstellar
medium and solar system determined only a lower bound $D/H \geq (1.6
\pm 0.2)\times 10^{-5}$.  In the standard three low-mass neutrino scenario
for Big Bang Nucleosynthesis (SBBN), this determines an upper bound tothe primordial mass
ratio,  $\eta_{10} \equiv 10^{10} n_B/n_{\gamma}
\leq (8.2 \pm 0.5)$ (\cite{Hata1997,Bludman}).
Recent observations (\cite{Tytler1996,Burles1996}) of 
deuterium absorption lines in two nearly primordial Lyman limit clouds 
illuminated by distant quasars (QSA) show $D/H=(2.4 \pm 0.3
\pm 0.3) \times 10^{-5}$ or $\eta_{10}=6.4^{+0.9}_{-0.7}$. 
 
Both these high values for $\eten$ are inconsistent with the low value
$\eten = 1.8 \pm 0.3$ once reported (\cite
{Carswell1994,Songaila1994,Rugers1996}) in light from quasar
Q0014+813, but since discredited (\cite{Tytler1997}).
This low $\eten$ had been in excellent concordance with
the $^4He$ abundance, 
\be 
Y_P=0.232 \pm 0.003 (stat) \pm 0.005 (syst)  
\ee
(\cite{Geiss,Olive}), formerly inferred from HII regions and the Standard
($N_\nu=3$ model) and in good agreement with the $^7Li$ Spite plateau
in halo stars (\cite{Spite}).  The higher $\eten$ value, on the other hand, was
inconsistent with the above $^4 He$ abundance and demanded $^7Li$
depletion in stars and surprisingly little Galactic chemical evolution
of $^2H,~^3He$. This discordance between the higher $\eten=6.4^{+0.9}_{-0.7}$ 
demanded by low D/H and the lower  $\eten=1.8 \pm 0.3$ demanded by the 
then-accepted Helium abundance (1) was the crisis for BBN (\cite{Hata1995}).

After this paper was first submitted for publication, improved measurements
of $D/H$ and $Y_p$ have ruled out the lower $\eta_{10}$ value and
ended the BBN crisis: (1) Improvements in the two QSA
determinations (\cite{Tytler1997}) now give $D/H=(3.2 \pm 0.4) \times
10^{-5}$, implying $\Ombe= 0.020 \pm 0.002, ~\eta_{10}=5.5 \pm 0.5, ~Y_p=0.247 \pm 0.0025$; (2)
This is consistent with $Y_p=0.243 \pm 0.003$ measured  (\cite{Izotov})
 in an important Wolf-Rayet galaxy I ZW 18 and in eight other
low-metallicity HII clouds.  These increases in $D/H$ and in $Y_p$
suggest concordance at values for $\eta_{10}$, $\Ombe, \Omega_B$ and
$\Omega_{cl}$ only 15\% lower than the values we adopted in the last
column of Table 1, sustaining our conclusions in Section 4.

\subsection{Implications for the Baryon and Total Mass Densities in the Universe}
 
Because the photon number density in the Universe is well determined
by the cosmic background temperature, the two incompatible values
$\eta_{10}=6.4 ^{+0.9}_{-0.7}$ or $(1.8 \pm 0.3)$, determine substantially
different baryon mass densities $\Omega_B=0.00366 \eta_{10} h^{-2}$,
where $h \approx 0.70 \pm 0.15$ is the present Hubble constant in
units of 100 km/sec/Mpc 
(\cite{Freedman1994,Riess1995,Ken}).
For the allowed range $H_0=(70 \pm 15)$ km/s/Mpc, either
$\Omega_B=(0.005-0.010)$ or $(0.019-0.030)$ (95\% C.L.)
(\cite{Hata1997}), which we hereafter we refer to as lower- and
higher-$\Omega_B$ (Table 1). Unless $h$ is large, the lower-$\Omega_B$
may already be excluded by present baryon inventories
(\cite{Fukugita1996,Bahcall1997}) showing $\Omega_Bh^2 \geq 0.03$.
In the next section, we discuss two different measures of the baryon
mass fraction $f_B \equiv M_B/M_{cl}$ in rich clusters of galaxies and
use them to calculate the total mass density in rich clusters,
$\Om \equiv \Omega_B/f_B$.  In Section III, we
calculate the limits on $\Omega_B/f_B$ for the lower- and
higher-$\Omega_B$ separately.  Because $f_B$ in rich clusters might be
enhanced over the cosmic value $\Omega_B/\Omega_m$ by a ``baryon
enhancement factor'' $\Upsilon$
(\cite{White1993,Steigman1995}), the cosmic total mass density $\Omm=
\Upsilon \Omega_B/f_B$ might differ from $\Om$ defined above.    

This paper extends  Hata {\em et al.} (1997) by including new data
(\cite{Evrard1997,Herbig1995,Myers})
 on the baryon fraction in rich clusters and new dynamical limits on
$\Omm$(\cite{Dekel1997}), and concludes that clusters are indeed a fair
sample of the Universe ($\Upsilon
\sim 1$), that  the baryon density is relatively high ($\Omega_B=(0.025-0.099)$
(95\% C.L.), that $h\approx 0.55$ allows a critical density Universe,
but that larger $h$ implies that we live in a medium density Universe.

\section{BARYON MASS FRACTION IN RICH CLUSTERS}

The luminous matter density $\Omega_{lum}=0.004 +0.007h^{-3/2}$
(\cite{Fukugita1996,Bahcall1997,Persic1992}).  The observed mean
$Ly \alpha$ flux decrement shows baryon density $\Omega_B \geq
0.0125h^{-2}$ (\cite{Weinberg1997}.  This shows that considerable
baryonic matter is dark and that SBBN predicts $D/H < 6 \times
10^{-5}$.  This already argues against the lower $\eten$ or $\Ombe$
choice

  In this section, we will summarize dynamical observations showing
that, the baryon fraction in rich clusters, $f_B <0.18$, so that rich
clusters are dominated by non-baryonic matter.  These rich clusters
are the largest virialized structures and, although of intermediate
size ($(1-10) h^{-1}~Mpc$) and constituting only a small fraction of
the total mass in the Universe, are thought to be fair samples of the
entire Universe.

\subsection{Baryon Fraction in Cluster Hot Gas}

The baryonic and total masses, $M_B,~M_{cl}$ in the hot gas in
clusters have each been measured by two different methods: originally
(\cite{White1991,White1995,Evrard1997}) from the X-ray bremmstrahlung
off hot cluster gas, and, more recently (\cite{Myers}), from the
Sunyaev-Zeldovich (SZ) inverse Compton spectral distortion of cosmic
background radiation.  Both methods depend on modeling the density and
thermal structure of clusters to determine the baryon mass in gas
$M_B$, the total mass $M_{\it cl}$, and, hence, the cluster baryon
mass fraction $f_B \equiv M_B/M_{\it cl} \equiv
\Omega_B/\Om$.

From detailed analysis of a large sample of rich clusters, and
comparison with CDM cluster models, Evrard (\cite{Evrard1997}) obtains
for the baryon fraction in hot gas
\be f_B(Xray,Evrard)= (0.060 \pm 0.003)~h^{-3/2}.  \ee 
This is consistent with 
\be f_B(Xray,select)=(0.054 \pm 0.013) ~h^{-3/2} \ee observed in
three large, well-studied  clusters A2142 (\cite{White1995}), A2256
(\cite{Henry}) and Coma (\cite{White1993}), that are 
probably rounder and smoother
than the other clusters in Evrard's larger, less selective catalogue.
(Both these X-ray measurements are consistent with earlier, less
sensitive measurements
(\cite{White1991}) \be f_B (Xray) =0.049^{+0.028}_{-0.014} ~h^{-3/2},
\ee which we do not show or use.)
 
From the Sunyaev-Zeldovich effect in these three rich clusters, Myers
{\it et al} (\cite{Myers}) obtain \be f_B (SZ)=(0.061 \pm
0.011)~h^{-1}. \ee Both values (2) and (4) are unweighted means, with
the errors combined in quadrature.  (Because a fourth cluster, A478,
has a SZ mass three times higher than these three and differs
significantly in other respects, we and Myers omit it from our
determination of $f_B$.  Had Myers included A478, he would have
obtained $ f_B (SZ)=(0.087 \pm 0.030)~h^{-1},$ insignificantly higher
than the baryon fractions (2) and (4).  Besides A478, we have also
omitted from the SZ analysis, clusters A665 (\cite{Birkenshaw}) and
CL0016+16, A773 (\cite{Carlstrom}) for which X-ray observations are
apparently not available.)

\subsection { Consistency of X-ray and SZ Determinations of Baryon
Fraction}

For these three selected clusters,
Myers {\it et al} obtain SZ hot gas masses that are on average $(1.27
\pm 0.13) h^{1/2}$ larger than the X-ray hot gas masses measured by
White and Fabian (\cite{White1995}), by Henry {\it et al} (\cite{Henry})  ,
and by White {\it et al} (\cite{White1993}), i.e. the X-ray and SZ hot gas
masses from these three nearby clusters are consistent at
\be f_B  (HG,select)=(0.12\pm 0.02)\ee
over the broad range $h=0.62^{+0.14}_{-0.12}$. For smaller $h$, this
selected cluster baryon fraction is some 11\% less than that observed
in Evrard's larger X-ray sample.  Although Myers excludes A478 from
his determination of $h$, when he includes it, he obtains
$h=0.54^{+0.12}_{-0.11}$, lower than but consistent 
with the value we obtained from the three selected clusters. 

\subsection{Baryon Fraction in Galaxies}

The baryons in hot gas must be augmented by the luminous mass
 $\Omega_{lum} \approx 0.009 $
 (\cite{Lynds1986,Soucail1987,Tyson1990,Fukugita1996,Bahcall1997}) and
 by dark baryons residing in galaxies.  Judging by the halo mass in
 our own Galaxy, the dark mass fraction could be as large as this
 luminous mass fraction, making $M_{gal}/M_{cl} = 0.0135 \pm 0.0045$.
 (Myers adopts the slightly smaller value $M_{gal}/M_{cl}= 0.009 \pm
 0.003$, observed in Coma (\cite{White1993})).  We therefore finallly
 adopt (Fig. 1) for the total baryon mass fraction, as measured in
 X-rays, \be f_B(Evrard)= (0.060 \pm 0.003)~h^{-3/2}+ 0.0135 \pm
 0.0045, \ee or \be f_B(select)=(0.054 \pm 0.013)~h^{-3/2}+ 0.0135 \pm
 0.0045 \ee and, as measured by the Sunyaev-Zeldovich effect,\be
 f_B(SZ)=(0.061 \pm 0.011)~h^{-1}+0.0135 \pm 0.0045. \ee The $1
 \sigma$ bounds on these three cluster total baryon mass fractions are
 shown in Figure 1 by solid and by dashed lines for Evrard's and for
 the three selected X-ray clusters respectively, and by the shaded
 area for the SZ observations.  For our preferred range
 $h=0.62^{+0.14}_{-0.12}$, $f_B \approx 0.13 \pm 0.03$.  For {\em
 larger h}, the three observations agree on $f_b \sim 0.09 \pm 0.02$.
 For {\em smaller h}, the larger values $f_b \sim 0.12 \pm 0.02$ from
 SZ and $f_b \sim 0.15 \pm 0.03$ from X-rays are suggested.  Note that
 including the X-ray measurements and the baryons in galaxies
 increases the baryon fraction over that in Eqn. (4).  In any case,
 the conservative bounds are $f_B \approx 0.11 \pm 0.03$ for larger
 $h$ and $f_B \approx 0.14 \pm 0.04$ for smaller $h$.

These upper and lower bounds on the baryon fraction provide lower and
upper limits on $\Omega_B/f_B \equiv \Om$, shown in Figure 2 for
lower-$\Omega_B$ (solid curves) and for higher-$\Omega_B$ (dashed
curves). The lower $\Omega_B h^2=(0.007 \pm 0.001)$, implies
$\Omega_{cl}< 0.27$.  The higher $\Omega_B h^2= (0.023 \pm 0.003)$
implies $\Om=(0.3-0.9)$.  This estimate is a little smaller than the value 
obtained (\cite{Myers}) using SZ data alone.  Nevertheless, $\Om$=1 is
still possible, for $h<0.6$.

We recall that baryon inventories (\cite{Fukugita1996,Bahcall1997}) 
already show $\Ombe \geq 0.03$ and that the cosmic virial
theorem already implies $\Omm \geq 0.2$.  If we were ready to accept
rich clusters as a fair sample of the Universe, this would already practically
exclude the lower $\Ombe$ value.  Nevertheless, in the next section,
we will test the possibility that the cluster and cosmic total mass
densities differ by a baryon enhancement factor $\Upsilon=\Omm/\Om$.

\section{DYNAMICAL MEASURES OF THE COSMIC MASS DENSITY $\Omega_{\lowercase{m}}$}

If $\Upsilon \sim 1$, the lower and higher $\Omega_B$ choices
 discussed in Section II would then already imply respectively a low
 and an intermediate or critical density Universe.  Although numerical
 simulations and other theoretical arguments
 (\cite{White1991,White1995}) strongly suggest that rich clusters do
 not appreciably concentrate baryons, we will now test this assumption
 by comparing $\Om$ from clusters with large-scale determinations of
 $\Omega_m$.  Most of these large-scale determinations depend on
 models for the evolution of large scale structure from assumed
 initial fluctuations and dark matter content.

The least model-dependent global bounds on $\Omm$ derive from (1)
diverging flows in voids, (2) from distant ($z \sim 0.4$) supernovae
Ia distance indicators, (3) from weak gravitational lensing of quasars
by intervening galaxies and rich clusters, and (4) from the expansion
age of the Universe, $t_0=H_0^{-1}f(\Omm,\Omega_{\Lambda})$.  From
diverging flows, $\Omm > 0.3~(2.4\sigma)$ because ``voids cannot be
more empty than empty'' (mass densities cannot be negative)
(\cite{Dekel1993}).  From type Ia supernovae (\cite{Perlmutter1997}),
$-0.3<\Omm-\Omega_{\Lambda}<2.5$ (90\% C.L.) or $\Omm>0.49$ (95\%
C.L.)  for a flat Universe.  From the mass-to-light ratios in weakly
lensed rich clusters (\cite{Kaiser}), $\Omega_m \simeq (0.3-1)$ and
from the statistics of gravitational lens counts (\cite{Kochanek}),
$\Omm > 0.34$ in a flat universe From the most likely age of the
oldest stars $t_0 \geq 12$ Gyr (\cite{Chaboyer,Reid}), implies the
upper bounds on $\Omm$ shown by the heavy dashed curves in Figure 2,
for an open cosmology with $\Omega_{\Lambda}=0$ and for a flat
cosmology with $\Omega_{\Lambda}=1-\Omm$.  If $\Upsilon=1$ and $t_0,
h,~f_B$ are all at their lower limits, a matter-closed universe
$\Omm=1$ is possible.
 
On large comoving scales $(10-100) h^{-1} Mpc$, the total mass density
derives from galaxy redshift surveys (subject to optical biasing),
from dynamical studies of cosmic flows, and from CBR growth of
fluctuations (free of optical biasing) .  Omitting the observations
which depend on optical biasing, Dekel et al (\cite{Dekel1997})
summarize: (1) From preliminary CAT and Saskatoon observations
(\cite{Hancock,Saskatoon}) of the first CBR acoustic peak,
$\Omm+\Omega_{\Lambda} > 0.3$ (95\% C.L.); (2) Cosmic Flows
(\cite{Zaroubi,Kolatt}) give $\Omm>0.3$ ; (3) Growth of Fluctuations
together with cluster morphology give $\Omm > 0.2$; (4 ) Cobe Power
Spectrum + Mark III Velocities gives model-dependent results: (a)
$\Omm \approx 0.2 h^{-1}$ from untilted CDM models, assuming no
optical biasing; (b) $\Omm \approx (0.45 \pm 0.07)$ from spatially
flat CDM with spectrum tilt $n$; (c) the best CDM fit requires a small
tilt and either $\Omm
\sim 0.7$ or $\Omega_{\nu} \sim 0.2$; (d) the first acoustic peak
requires a small tilt and a high baryon content, $\Omega_B \sim 0.1$
(\cite{Zaroubi}).

All these observations of large-scale structure require $\Omm > 0.3$,
so that since $f_B=0.14 \pm 0.04~(0.11 \pm 0.03)$ for small (large) $h$, we have
$\Upsilon \Omega_B > 0.03~(0.024)$.  Unless rich clusters are
baryon-enhanced by the unreasonable factor $\Upsilon \sim 3$, 
this rules out the lower-$\Omega_B$
solution, $\Omega_B<0.01$ (95\% C.L.).  The large-scale structure
observations are consistent with the higher-$\Omega_B$ observations,
$\Omega_B=(0.025-0.099)$ (95\% C.L.)  and with rich clusters being a
fair sample of the Universe.  If $h$ is near its lower bound, and
$\Omega_B$ near its upper bound, a critical density universe
$\Omega_m=1$ is just allowed.

\section{CONCLUSIONS}

We now summarize (Table 1) our
cosmological conclusions, distinguishing the implications of the
baryon fraction observed in X-ray and SZ studies of rich clusters, from
the implications of large-scale structure. In all cases, we assume
$H_0=(55-85)~ $ km/s/Mpc.

The lower value $\Omega_B=(0.008-0.033)$ (95\% C.L.) together with
the baryon fraction in rich clusters would have implied a low
cluster mass density $\Omega_{cl} < 0.27$,  inconsistent with baryon
inventories and with rich clusters being a fair sample of the
Universe.

The larger value $\Omega_B=(0.025-0.099)$, together with the baryon
fraction observed in both X-ray and SZ clusters, allows a higher
matter density $\Omega_{cl}=(0.3-0.9)$, consistent with rich clusters
being a fair sample of the Universe and with data from large-scale
structure.  If $\Omega_B$ is near its upper bound and $h$ near its
lower bound, a critical density is possible.

This work is supported by Department of Energy Contract
No. DE-AC02-76-ERO-3071 and has benefited by spirited
discussions with G. Steigman.

\renewcommand{\baselinestretch}{1.3}
%#ifdef PREPRINT
%\renewcommand{\baselinestretch}{1.0}
%#endif

\vspace{7ex}
\begin{table}[hbt]
\caption{
Cosmological implications of lower-$\Omega_B$ (consistent with previously
observed $^4 He, ~^7 Li$ abundances) and of higher-$\Omega_B$
(consistent with the deuterium abundances observed in the nearby Galaxy and in
two quasar aborption systems
(\protect{\cite{Tytler1996,Burles1996,Geiss,Hata1996}}).
The errors are for 68\% C.L., while the 
ranges in the parentheses are for 95\% 
C.L. We take $H_0=(70 \pm 15)~km/s/Mpc$, so that $h^2=(0.49 \pm 0.21)$.
In each case, the mass density in clusters, $\Omega_{cl} \equiv \Omega_B/f_B$, 
is obtained from the range in $f_B$ in Figure 1.
(If the latest determinations
(\protect{\cite{Tytler1997}}) $D/H=(3.2 \pm 0.4)10^{-5},~\eta_{10}=5.5\pm 0.5$
were now used, $\Ombe , \Omega_B , \Omega_{cl}$ in the last column and in 
Fig. 2 would now be reduced by 15\%.)   
}

\label{tab:constraints}
\vspace{6.0ex}
\begin{tabular}{l  c c }
\hline%-----------------------------------------------------------------------
\hline%-----------------------------------------------------------------------
		& lower-$\Omega_B$
                & higher-$\Omega_B$  \\
%\hline%-----------------------------------------------------------------------
%\hline%-----------------------------------------------------------------------

\ D/H (10$^{-5}$) 
                & $19 \pm 4$  
                & $2.4 \pm 0.3 \pm 0.3$, $\ge 1.6 \pm 0.2$ \\
\hline%--------------------------------------------------------------
$\eta_{10}$     & $1.8 \pm 0.3$ (1.7 -- 2.7)   
                & $6.4^{+0.9}_{-0.7}$ (5.1 -- 8.2)\\
               
$\Omega_{\rm B}h^2$
                & $0.007 \pm0.001$ 
                & $0.023 \pm 0.003$ \\
                & (0.005 -- 0.010)
                & (0.019 -- 0.030)\\

$\Omega_B$    
	& (0.008 --0.033)
	& (0.025 --0.099) \\
\hline%---------------------------------------------------------------
$\Omega_{cl}$ 
	& 0.09 --0.26
	 &0.3 -- 0.9\\
 %
%\hline%-----------------------------------------------------------------------
%\hline%-----------------------------------------------------------------------
%
\end{tabular}

\end{table}

%****  Figures Captions****************************************************************

%                    *            *           * 
%
%				Figure 1
%
\begin{figure}[h]

\caption{
The allowed $1\sigma$ range of baryon fraction $f_B$ in rich clusters of 
galaxies as function of the Hubble constant 
$H_0=100h$ km/s/Mpc, as measured by thermal bremmstrahlung X-rays
(in Evrard's catalogue(\protect{\cite{Evrard1997}})(thin solid curve)
and in three selected rich clusters 
(\protect{\cite{White1995,Henry,White1993}})(dashed curve))
and by the Sunyaev-Zeldovich upscattering of cosmic background
radiation (shaded region SZ) (\protect{\cite{Myers}}). In all three cases,
the baryon fraction measured in hot gas has been augmented by the luminous 
and dark baryons in galaxies and stars.  }
%
% ~/Graphics/SZXaugment
\end{figure}
   
%			*	*	*
%

%				Figure 2
%
\begin{figure}[h]

\caption{
BBN and cluster baryon fraction constraints on the clustered matter
density $\Omega_{cl} \equiv \Omega_B/f_B$ as function of Hubble
constant $h$ for $\Ombe=0.007 \pm 0.001$ (solid curves) and for
$\Ombe=0.023 \pm 0.003$ (dashed curves).  The regions between each
pair of curves are allowed at 68\%(95\%) C.L.  The heavy dashed
curves are upper bounds on the {\em cosmic} baryon density
$\Omm$ in open and flat Universes of age
greater than $12~ Gyr$. 
In principle, $\Omm$ might differ from the clustered baryon
density $\Om$ by a factor $\Upsilon \equiv \Omm/\Om$.  In fact, observation
of large-scale structure show $\Omm \sim \Om$, so that $\Upsilon \sim
1$: rich clusters are a fair sample of the Universe. 
%\label{clusteredmass}   
}

\end{figure}

\psfig{figure=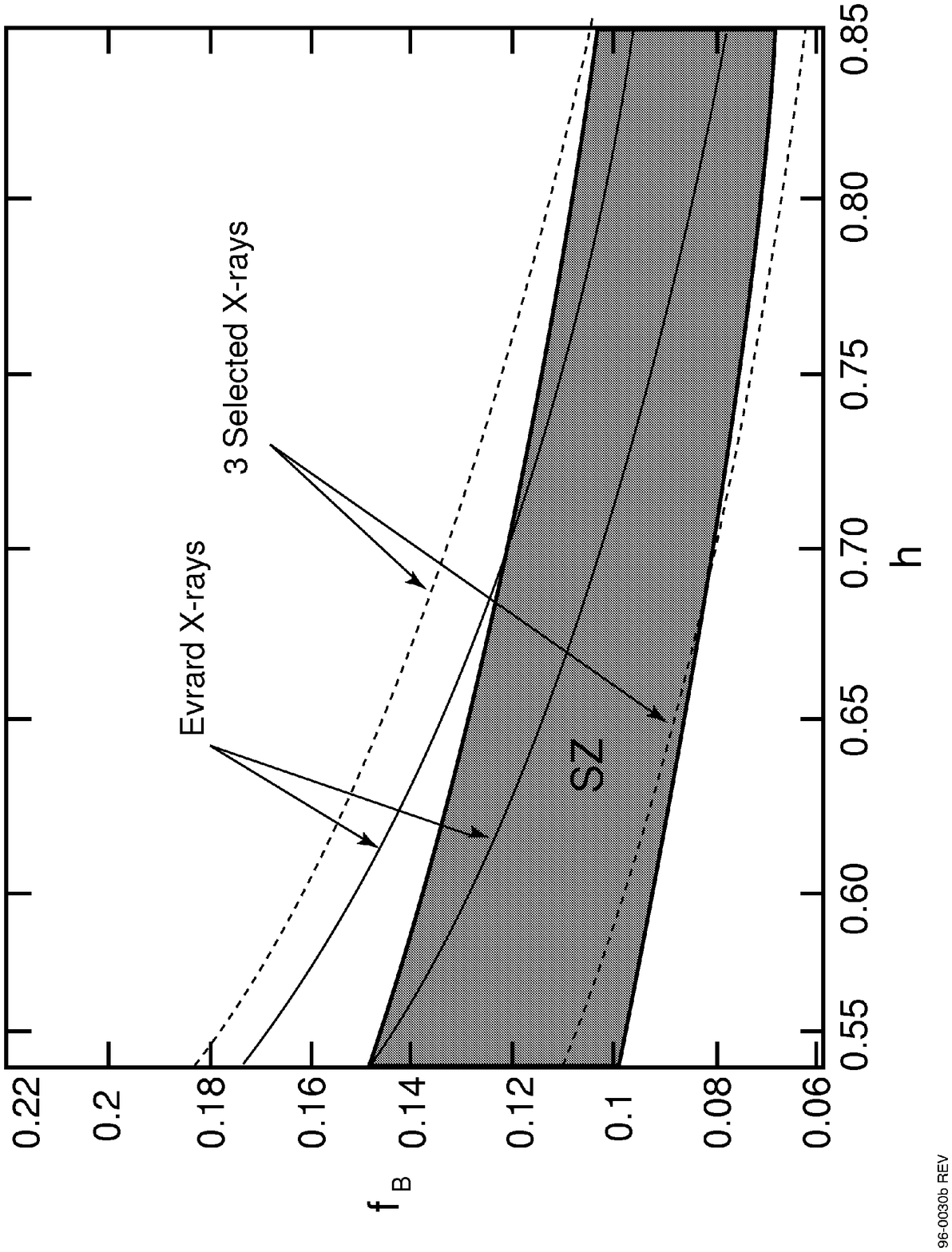,height=\textheight}
\psfig{figure=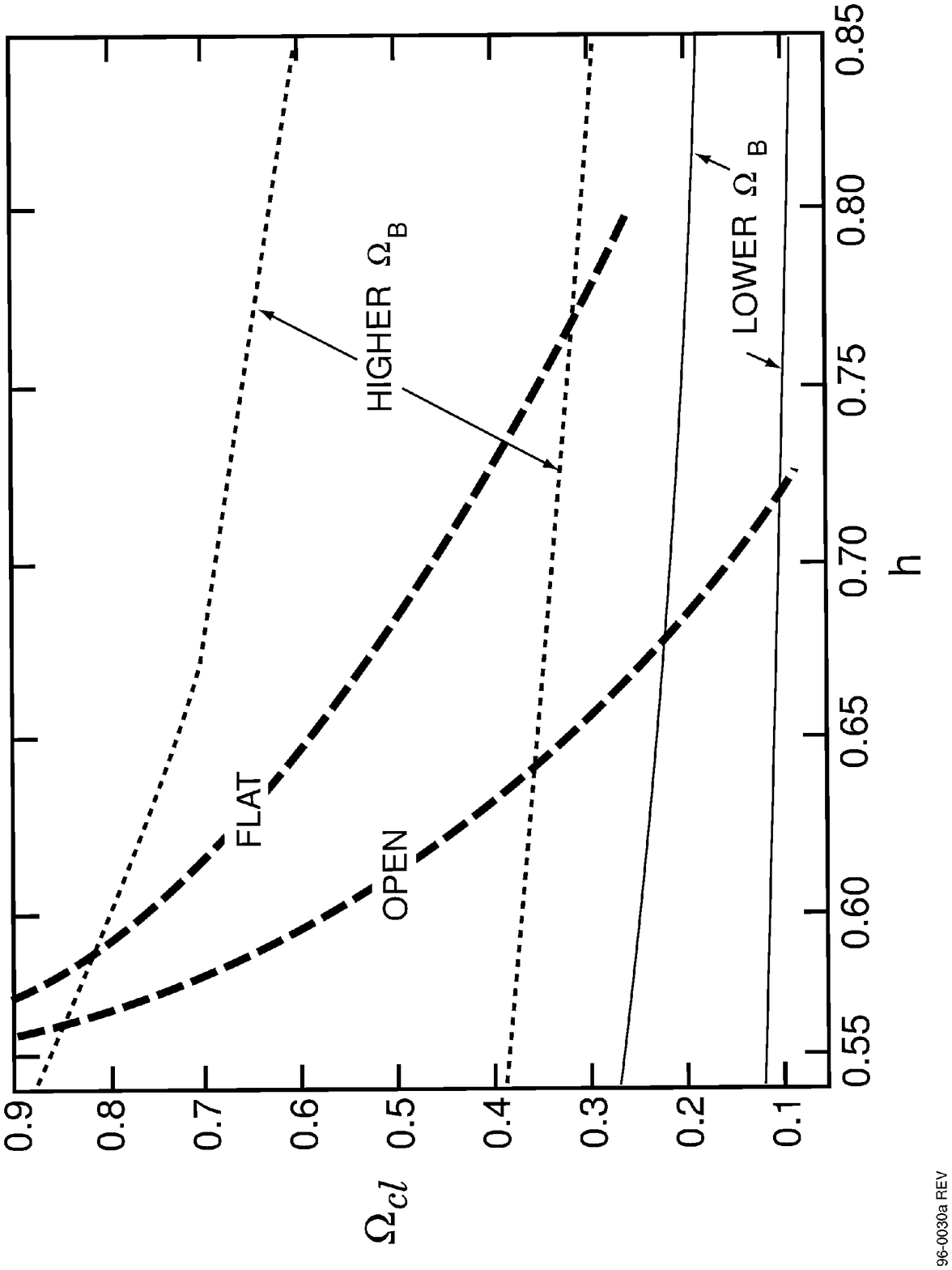,height=\textheight}

\end{document}